\documentstyle[preprint,aps]{revtex} 
\begin {document} 
\draft 
 
%\wideabs{ 
\title{\bf Quasi one dimensional $^4$He inside carbon nanotubes}   
\author{M.C. Gordillo, J. Boronat and J. Casulleras}  
\address{ 
Departament de F\'{\i}sica i Enginyeria Nuclear, Campus Nord B4-B5,  
Universitat Polit\`ecnica de Catalunya. E-08034 Barcelona, Spain}  
\date{\today} 
 
\maketitle 
\begin{abstract} 
We report results of diffusion Monte Carlo calculations for both $^4$He 
absorbed in a narrow single walled  
carbon nanotube ($R$ = 3.42 \AA) and 
strictly one dimensional $^4$He.  
Inside the tube, the binding energy of liquid $^4$He is approximately three 
times larger than on planar graphite. At low linear densities, $^4$He in a 
nanotube is an experimental realization of a one-dimensional quantum 
fluid. However, when the density increases the structural and energetic 
properties of both systems differ. At high density, a 
quasi-continuous liquid-solid phase transition is observed in both cases. 
\end{abstract} 
 
%} 
 
Since their discovery by Ijima \cite{Ijima} in 1991,  carbon nanotubes  
have received a great deal of attention. Basically, they are the result of 
the seamless rolling up of one or several graphite sheets over themselves 
\cite{iji2,bethune,thess}. 
Depending on the relative  
orientation of the rolling axis with respect underlying graphite structure, one  
can have different types of nanotubes \cite{ham}: armchair, zig-zag and chiral  
with different radii and different mechanical and electrical properties.  
Nowadays, it is 
possible to obtain high yields of nanotubes (single and multiple walled), 
with a variety of diameters ranging from 7 to 40 \AA \cite{3} and  
lengths up to $\sim$ 1000 times larger. 
 
One of the most attractive features of carbon nanotubes is the possibility 
of filling with different materials both their inner cavities and the interstitial 
channels among them \cite{pen,ajayan}. The interest in this field is twofold. 
On one hand, the  
expected increase in the particle-substrate potential energy with respect to a flat 
carbon surface has suggested the use of nanotubes as storage devices  
for molecular hydrogen in fuel cells \cite{dillon,levesque}. On the other, more theoretical, nanotubes 
provide a reliable realization of one-dimensional systems in the same way that  
a substance adsorbed on graphite manifests trends that are characteristic of a  
two-dimensional medium. If the nanotubes are filled with light atoms (He) or molecules 
(H$_2$) and the temperature is low enough, one is dealing with quasi-one dimensional 
quantum fluids. Such an experimental realization has been carried out for the first 
time by Yano {\em et al.} \cite{yano} in a honeycomb of FSM-16. This is a mesoporous substrate with 
tubes approximately 18 \AA$ $ in diameter. Using a torsional oscillator, this group 
proved the existence of superfluidity of the $^4$He atoms absorbed in the pores 
below a critical temperature of $\sim$ 0.7 K. More recently, Teizer {\em et al} 
\cite{tei} have 
studied experimentally the desorption of $^4$He previously absorbed in the interstitial 
sites of carbon nanotube bundles. In this case, the data points unambiguously to the 
one-dimensional nature of the helium inside the nanotubes.  
 
From a theoretical point of view, it has been recently established using 
both the hypernetted chain (HNC) variational approach   
\cite{krot} and the DMC method \cite{stan3} that 
strictly one dimensional (1D) $^4$He is a self-bound liquid at zero  
temperature. However, contrary to the situation for dilute classical gases   
\cite{stan3,stan1,stan2}, there are no many body calculations of 
quantum fluids inside nanotubes yet.   
In this work, we address the question of the quasi-one dimensionality 
of $^4$He absorbed in a tube by a direct comparison between the 
results of diffusion Monte Carlo (DMC) for strictly 1D $^4$He and $^4$He  
inside a nanotube of radius equal to 3.42 \AA, which corresponds to a (5,5) 
armchair tube in the standard nomenclature \cite{iji2}.  
 
The DMC method \cite{rey,boro} solves stochastically the $N$-body  
Schr\"odinger equation 
giving results that are {\em exact} for bosonic systems as liquid $^4$He,  
provided that the interatomic potential is known. In the present 
calculation, we have used the HFD-B(HE) Aziz potential for the He-He pair interaction 
\cite{aziz1}, and the potential given by Stan and Cole \cite{stan1} in their 
study of Lennard-Jones fluids in tubes for the  
He-tube one.  Basically, they consider  
the nanotubes as smooth cylinders by making a $z$-average of the corresponding       
sum of all the C-He interactions. Thus, the potential felt by a particle 
only depends on its distance to the center of the cylinder. This is a 
simplification, but one would expect the error involved to be small  
since the helium atoms are much larger than the C-C distance.  
In fact, the differences in energy and position between a $^4$He atom 
 in the smooth  
cylinder model and the same particle considering its interaction with the  
surrounding individual carbons are about 1\% for the tube considered  
here \cite{jcp}. 
 
The efficiency of the DMC method is greatly enhanced by introducing a trial wave 
function $\Psi({\bf R})$ that acts as an importance sampling auxiliary function. 
In 1D $^4$He we have used a two-body Jastrow wave function 
\begin{equation} 
\Psi^{\rm 1D}({\bf R}) = \Psi_{\rm J}({\bf R}) 
\end{equation} 
with $\Psi_{\rm J}({\bf R}) = \prod_{i<j} \exp \left[ -\frac{1}{2} \left 
(\frac{b}{r_{ij}} \right)^5 
\right]$, whereas liquid $^4$He inside nanotubes requires the additional introduction 
of a one-body term 
\begin{equation} 
\Psi^{\rm T}({\bf R}) = \Psi_{\rm J}({\bf R}) \Psi_{\rm c}({\bf R}) 
\end{equation}   
with $\Psi_{\rm c}({\bf R}) = \prod_i^N \exp(-c \, r_i^2)$ ($r_i$ being the radial distance of the 
particle to the center), that accounts for the hard core of the helium-nanotube interaction. 
Using the variational Monte Carlo method (VMC) we have optimized the parameters 
$b$ and $c$ at low densities around the equilibrium. The values obtained,  
$b$ = 3.067 \AA$ $  
and $c$= 2.679 \AA$^{-2}$, show a negligible dependence with the density 
and therefore they have been used everywhere in the DMC calculations. In
all the simulations $N=30$ atoms have been used, a number that has proved
to be large enough to reduce the size effects to the level of the
statistical errors reported.

The variational HNC equation of  Krotscheck and Miller \cite{krot} on 1D liquid $^4$He points to 
the existence of a liquid-solid phase transition at high density. We have explored 
this feature, that is only possible at zero temperature, by using a solid trial  
function which results from the product of $\Psi^{\rm 1D}({\bf R})$ and 
$\Psi^{\rm T}({\bf R})$ by a 
$z$-localized factor $\Psi_{\rm s}(R) = \prod_i^N \exp( -a(z_i- z_{is})^2)$. The solid sites 
$z_{is}$ are equally-spaced points along the $z$ direction which is both the longitudinal 
axes of the tube and the line of the 1D system. In both solid systems, a VMC optimization 
at high densities leads to values $b$= 2.939 \AA$ $ and $a$ = 0.612 
\AA$^{-2}$, with   
$c$= 2.908 \AA$^{-2}$ in the tube case.     
 
The energy per helium atom versus the linear coverage, 
$\lambda$, for the 1D (open squares, energy scale on the right) and the tube 
(full squares, energy scale on the left) is shown in Fig. 1. The two curves have have been 
drawn for the full square with the lowest $\lambda$ to coincide with the 
open  symbol for the same He density. One observes that for $\lambda <$  
0.05 \AA$^{-1}$, both curves are similar, 
but for larger concentrations the tube curve is located below the other one. 
A similar phenomenology appears in the comparison between the energies of purely 
2D $^4$He and $^4$He adsorbed in graphite. As for this system, the difference in energy between  
$^4$He in a nanotube and 1D $^4$He is always negative with an absolute value that  
increases with the density. Both in graphite and in nanotubes this increase  
with respect to the 2D and 1D systems is mainly due 
to the emergence of their actual 3D nature.  
Beyond this qualitative 
agreement between $^4$He adsorbed on graphite and inside carbon nanotubes, there 
are significant differences in the values of the binding energies in the two systems.  
The binding energy of a single $^4$He atom in graphite is $E_B^G$= 140.74 K 
\cite{whit}, whereas in the 
nanotube we are studying is roughly three times larger, $E_B^T$ = 429.97 K, a significant 
difference that has been dramatically observed in the desorption experiment of 
Teizel {\em et al} \cite{tei}. On the other hand, the departure of the real 
3D  systems ($^4$He in graphite or a nanotube) from the idealized 2D or 
1D liquids can be quantified by means of the parameter 
\begin{equation} 
\Delta^{\rm T(G)} =  \frac{(E^{\rm T(G)} - E_{\rm B}^{\rm T(G)})- E_{\rm 1(2)D}}  
{(E^{\rm T(G)} -  E_{\rm B}^{\rm T(G)})}  
\end{equation} 
where T stands for the tube and G for the graphite adsorbents and $E$ is the 
energy per particle in the system under consideration.  
Around the respective equilibrium densities one obtains  
$\Delta^{\rm T}$ = 90 \% and 
$\Delta^{\rm G}$ = 6 \%, a large difference that indicates that the 1D representation of 
$^4$He inside the nanotube is worse than the 2D modelization of  
$^4$He in planar graphite.   
 
Up to $\lambda$ = 0.15 \AA$^{-1}$ the energies per particle ($e=E/N$) of both 1D  
$^4$He and $^4$He inside the tube may be well fitted by a third-degree  
polynomial 
\begin{equation} 
e = e_0 +  A \left( \frac{ \lambda - \lambda_0 }{\lambda_0} \right)^2 + 
B \left( \frac{ \lambda - \lambda_0 }{\lambda_0} \right)^3 \ . 
\end{equation} 
The optimal values for the parameters $A$, $B$, $\lambda_0$, and $e_0$ are 
reported in Table I. The linear equilibrium densities $\lambda_0$ of both 
systems are close,  $\lambda_0^{\rm 1D}=0.062$ \AA$^{-1}$ and 
$\lambda_0^{\rm T}=0.079$ \AA$^{-1}$, whereas the energy difference between 
$\lambda$ = 0 and $\lambda=\lambda_0$ is significantly different, 0.00364 
and 0.018 K for 1D $^4$He and $^4$He inside the tube, respectively. The 
latter results point again to a large enhancement of the binding energy of 
$^4$He inside the tube with respect to the 1D system. On the other hand, 
the equilibrium density of liquid $^4$He inside the tube ($\rho_0=0.0022$ 
\AA$^{-3}$) is much smaller than the one in homogeneous 3D liquid $^4$He
($\rho_0 = 0.022$ \AA$^{-3}$).

In agreement with the DMC calculation of Stan {\em et al.} \cite{stan3} and 
the variational one  of  Krotscheck and Miller \cite{krot}, $^4$He selfbounds 
in a 1D array but with a binding energy (-0.0036$\pm$0.0002 K) much smaller
than that in 
2D (-0.897$\pm$0.002 K) \cite{stefi} and 3D (-7.267$\pm$0.013 K) \cite{boro}. It is worth noting that 
such a small total energy results from a big cancellation between the 
potential and kinetic energies. At $\lambda_0$, we have T/N = 
0.2706$\pm$0.0004 K and 
V/N = -0.2742$\pm$0.0004 K. In fact, the influence of the $^4$He 
interatomic potential in this system is very large.  A calculation at the equilibrium 
density $\lambda_0$ for the 1D system using the HFDHE2 Aziz potential \cite{aziz2} 
indicates that $^4$He is still a liquid, but the total energy is a factor 
 two smaller (-0.0018 $\pm$ 0.0003 K, with a potential energy 
 -0.2724$\pm$0.0004 K 
and the same kinetic energy). This sizeable differences   partially 
explain the discrepancies of the DMC calculation of Stan {\em et al.} 
\cite{stan3}  and 
ours with the results of Krotscheck and Miller \cite{krot} who used the HFDHE2 Aziz 
potential.   
 
From the values of the energy, one can obtain the linear system pressure, 
$p_{\lambda} = \lambda^2 \partial e/ \partial \lambda$, and estimate the 
same property for helium inside 
the cylinder as $p = p_{\lambda}/\pi R^2$. 
Fig. 2 displays this observable as a function of the $^4$He density.  
The  range  
corresponds to a liquid structure both in the 1D and tube cases (see 
discussion below). One can see that  
the pressure increases faster in a pure linear arrangement of atoms than in a tube. 
This can be understood if one considers that in a narrow nanotube it is 
possible to avoid the repulsive core of the nearest neighbors by shifting 
transversely the helium positions, a situation that is 
obviously not possible in 1D. 
Also interesting is the comparison  
between the sound velocity, $c(\lambda) = 
[1/m (\partial P / \partial \lambda)]^{1/2}$ at their respective equilibrium  
densities. The values are $c_{\rm 1D}$ = 
7.98 $\pm$ 0.07 m/s and $c_{\rm T}$ = 14.2 $\pm$ 0.8 m/s, in both cases a tiny 
fraction of the corresponding 2D ($c$ = 92.8 m/s) \cite{stefi} and 3D (238.3 m/s) 
\cite{boro} $^4$He liquids.  
The spinodal point can be obtained as the density at  which the speed of 
sound becomes zero. According to our results the  
spinodal points are located at $\lambda_{\rm 1D}$ = 0.047 
$\pm$ 0.001 \AA$^{-1}$ and $\lambda_{\rm T}$ = 0.059 $\pm$ 0.001 \AA$^{-1}$. 
 
Another aspect that has deserved our attention has been the existence of a  
liquid-solid phase transition at high densities. Evidences of 
this phase transition, that is only possible at zero temperature, appear in a 
variational calculation of 1D $^4$He \cite{krot}. A comparison between the DMC energies for  
the liquid and solid phases is given in Table II.    
One can see that in  
both systems, the energy per particle when localization 
is imposed ($a \neq 0$) is below the corresponding to a liquid structure
($a = 0$) for lineal 
densities greater than 0.358 \AA$^{-1}$. By means of the Maxwell double tangent 
construction, 
one would be able in principle to tell the solid from the liquid 
and to obtain the freezing and melting densities. Unfortunately, the energy 
differences between the $z$-localized and the liquid structures are too 
small to allow us to carry out a meaningful calculation. Our results indicate  
that for large enough densities ($\lambda >$ 0.358 \AA$^{-1}$) both  
the 1D and the tube arrangements 
have a localized (solid) phase and that the discontinuity in the density  (if any)  
is surely very small. This is in agreement with the results  
discussed by Withlock {\em et al.} \cite{whit} about the reduction of the 
size of this discontinuity from three to two dimensions: in 3D   
is fairly large, being considerably smaller in a 
purely 2D system. In 1D, we observe a further reduction towards  a 
continuous or a quasi-continuous transition. It is also remarkable that  
$^4$He inside the carbon tube remains a liquid up to a much larger pressure 
(around 5 times) than in bulk liquid $^4$He($\sim$ 2.6 MPa).

Information on the spatial distribution of the $^4$He atoms may be drawn 
from the two-body radial distribution function along the $z$ direction, 
$g_z(r)$. The functions $g_z(r)$ for 1D $^4$He and $^4$He in the tube are 
shown in Fig. 3 at several linear densities. Near the equilibrium density 
($\lambda$ = 0.08 \AA$^{-1}$, lower part of the figure) $g_z^{\rm 1D}(r)$ 
is quite similar to $g_z^{\rm T}(r)$, as corresponds to a quasi-one dimensional 
system. The same could be said in a broad range of densities, as it can be 
seen in the curves for $\lambda$ = 0.182 \AA$^{-1}$ (middle part of the 
figure). On the other hand, in the solid phase ($\lambda$ = 0.406 \AA$^{-1}$), 
$g_z^{\rm 1D}(r)$ and $g_z^{\rm T}(r)$ are different:   
in this case, the 3D nature of $^4$He inside 
the tubes produces a significant decrease in the localization with respect 
to the 1D result.  
 
In conclusion, we have compared the properties of strictly 1D $^4$He with   
$^4$He inside a narrow carbon nanotube using the diffusion Monte Carlo method.  
For a wide range of densities, $^4$He is a liquid in both systems, and also  
in both cases a quasi-continuous liquid-solid phase transition has been observed. 
In accordance with recent experimental determinations, the present calculation 
evidences a quasi-one dimensional behaviour of $^4$He inside a nanotube but 
significant differences with the ideal 1D system appear, specially when the  
linear density is increased. The origin of these differences is mainly the
existence of the additional transverse degree of freedom that helium atoms 
have inside a nanotube.
 
One of us (M.C.G.) thanks the Spanish Ministry of Education and Culture 
(MEC) for a postgraduate contract. This work has been partially supported 
by DGES (Spain) Grant No. PB96-0170-C03-02.   
 
\begin{table} 
\begin{tabular}{lcc}  
Parameter   &  1D $^4$He  & $^4$He in a tube \\ \hline 
$\lambda_0$ (\AA$^{-1})$ & 0.062 $\pm$  0.001 & 0.079 $\pm$ 0.003 \\ 
$e_0$ (K)               & -0.0036 $\pm$ 0.0002 & -429.984 $\pm$ 0.001 \\ 
$A$   (K)               &  0.0156 $\pm$ 0.0009 & 0.048 $\pm$ 0.006 \\ 
$B$   (K)               &  0.0121 $\pm$ 0.0008 & 0.0296 $\pm$ 0.009 \\ 
$\chi^2/\nu$            &         2.2         &    0.24           \\   
\end{tabular} 
\caption{Parameters of Eq. 4 for the two systems studied.} 
\end{table} 
 
\begin{table} 
\begin{tabular}{lcccc} 
$\lambda$ (\AA$^{-1}$) &  $E/N$ (1D, $a$ = 0)  & $E/N$ (1D, $a \not=$ 0) &  
$E/N$ (T, $a$ = 0) & $E/N$ (T, $a \not=$ 0) \\ \hline   
0.406  &  123.726 $\pm$    0.012  &  123.561  $\pm$  0.012 &  -350.155 $\pm$ 0.030 & -350.20 $\pm$ 0.02 \\ 
0.380  &   67.070 $\pm$    0.011  &   67.000  $\pm$  0.009 &  -382.282 $\pm$ 0.016 &  -382.321 $\pm$ 0.012 \\ 
0.358  &   37.602 $\pm$    0.008  &   37.596  $\pm$  0.007 &  -401.873 $\pm$ 0.013 & -401.844 $\pm$ 0.010 \\ 
0.338  &   21.881 $\pm$    0.007  &   21.904  $\pm$  0.005 &  -413.091 $\pm$ 0.014 &  -413.061 $\pm$ 0.012 \\ 
0.320  &   13.240 $\pm$    0.005  &   13.258  $\pm$  0.006 &  -419.551 $\pm$ 0.011 &  -419.493 $\pm$ 0.010 
\end{tabular} 
\caption{Energies per particle at large $\lambda$ for the systems 
studied. All the energies are  in K. See text for further details} 
\end{table} 
\noindent 
Figure captions  
 
1. Energy per particle ($E/N$) versus the linear concentration ($\lambda$), 
for the two  systems we have studied: a strictly one dimensional system 
(open squares, right energy scale), and a (5,5) armchair tube (full squares, 
left energy scale). In the first case, the error bars are less than the size 
of the symbols. Both energy scales are in K.  
 
2. Pressure at high helium densities for both systems (one dimensional,  
dashed line, left scale; nanotube, full line, right scale). 
 
3. Pair distribution function along the  
$z$ coordinate, $g_z^{\rm T(1D)}(r)$. Full lines correspond to the narrow tube at 0.08 \AA$^{-1}$ 
(bottom), 0.182 \AA$^{-1}$ (middle) and 0.406 \AA$^{-1}$ (top), and dashed lines to the purely 
linear system at the same densities.

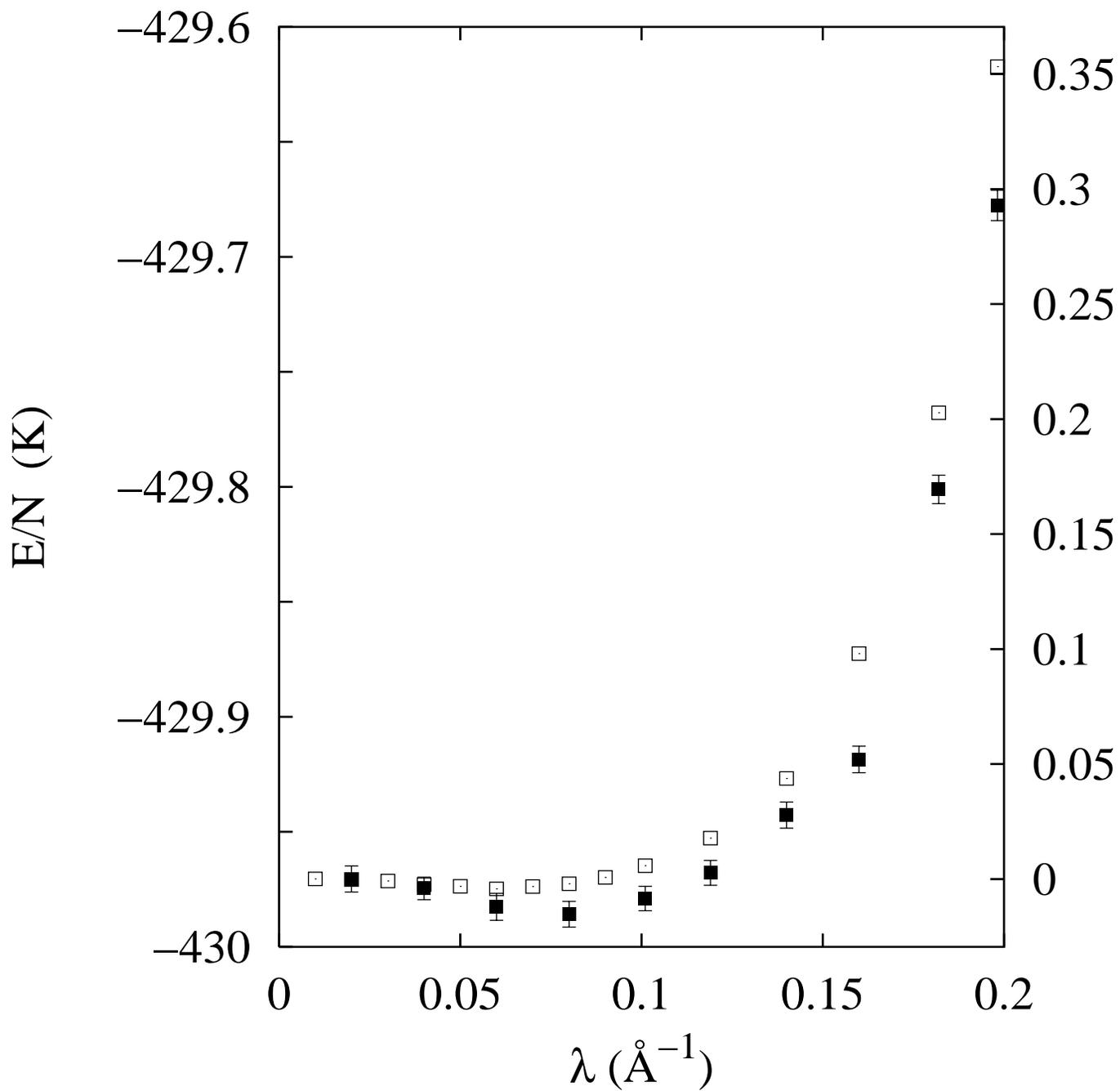
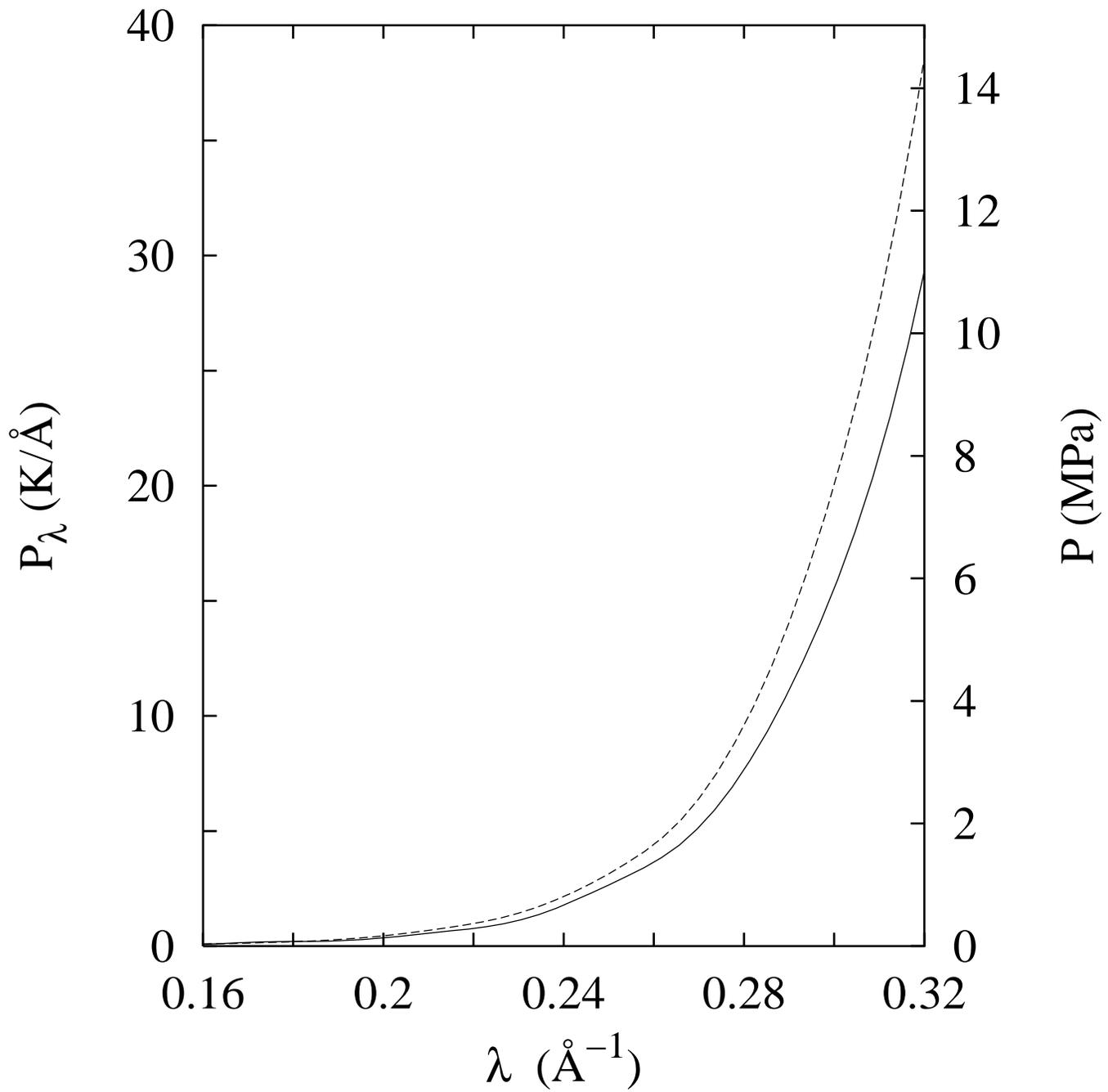
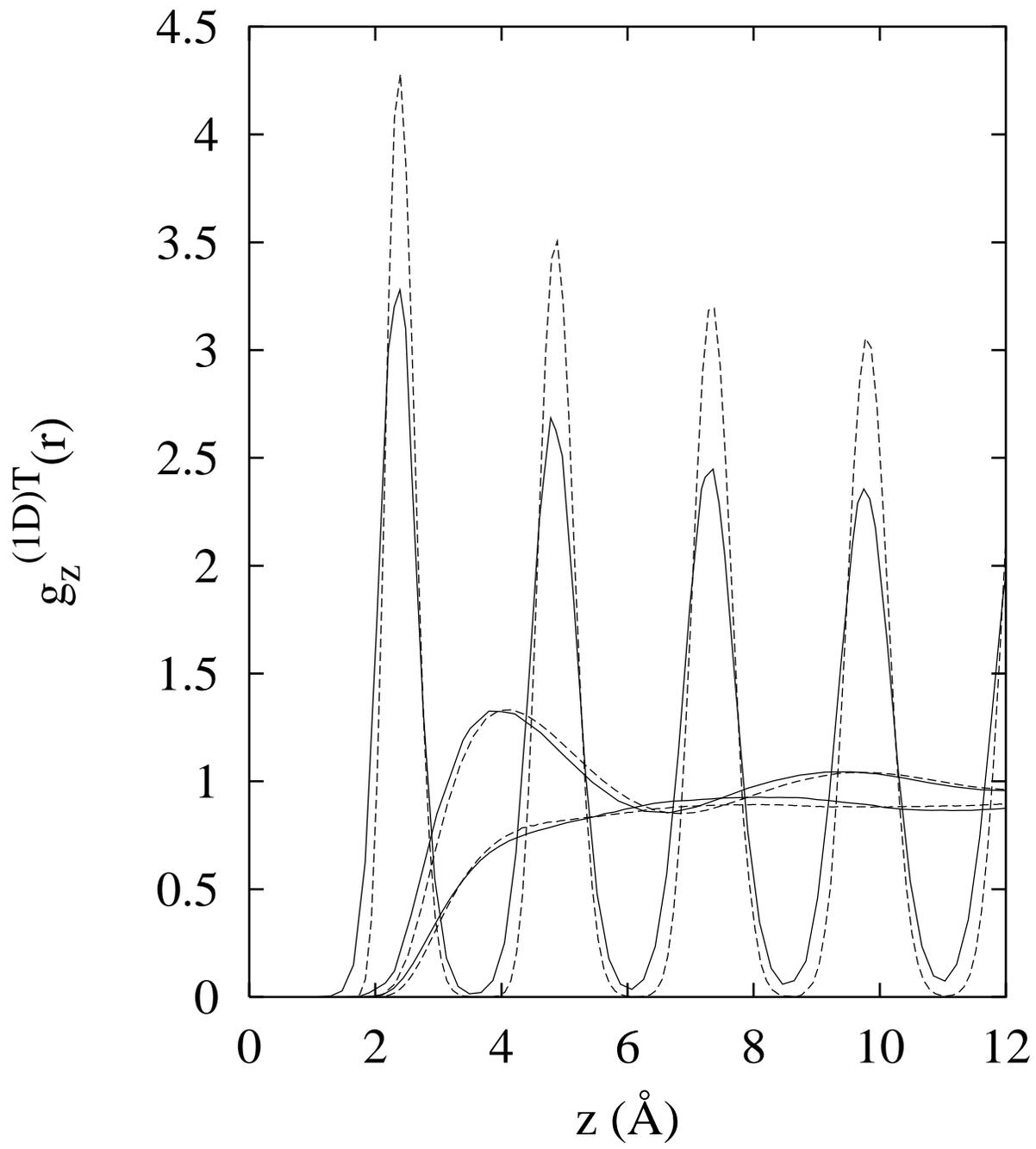
\end{document}